\documentclass[12pt,a4paper]{article}
\usepackage[utf8]{inputenc}
\usepackage{graphicx}
\usepackage{times}
\usepackage{color}

\usepackage{amsmath,amssymb,amsthm,graphicx,color,bm,soul}

\begin{document}

\title{Quantum-like model for unconscious-conscious interaction and emotional coloring of perceptions and other conscious experiences} 
\author{Andrei Khrennikov\\ 
Linnaeus University, International Center for Mathematical Modeling\\  in Physics and Cognitive Sciences
 V\"axj\"o, SE-351 95, Sweden}

\date{}                     

\maketitle

\abstract{Quantum measurement theory  is applied to quantum-like modeling of coherent generation of perceptions and emotions and generally for emotional coloring of conscious experiences. In quantum theory, a system should be separated from an observer.  The brain performs self-measurements. To model them, we split the brain into two subsystems, unconsciousness  and consciousness. They correspond to a system and an observer. The states of perceptions and emotions are described through the tensor product decomposition of the 
unconscious state space; similarly, there are two classes of observables,  for conscious experiencing of perceptions and emotions, respectively. Emotional coloring is coupled to quantum contextuality: emotional observables determine contexts.  Such contextualization reduces degeneration of unconscious states. The quantum-like approach should be  distinguished from consideration of the genuine quantum physical processes in the brain (cf. Penrose and Hameroff). In our approach the brain is a macroscopic system  which information processing can be described by the formalism  of quantum theory.}, 

{\bf keywords:} Quantum-like model, unconscious-conscious interaction, emotional coloring, perceptions, quantum measurement theory

\section{Introduction}

We start the paper with citation from \cite{Paper}: \footnotesize{{\it ``Although emotions, or feelings, are the most significant events in our lives, there has been relatively little contact between theories of emotion and emerging theories of consciousness in cognitive science.''}} We want to formalize this contact  in the quantum-like framework for cognition by generalizing the sensation-perception model based on quantum information theory \cite{KHRfrontiers}.

In the present paper, the quantum theory of measurement is applied to quantum-like modeling of coherent generation of perceptions and emotions and generally for emotional coloring of conscious experiences. In quantum theory, a system should be separated from an observer 
(or a measurement apparatus) \cite{BR}. In the quantum framework, this separation is very important  -- although establishing the sharp boundary between a system and a measuring apparatus  is
a difficult problem.\footnote{It becomes even more difficult if a conscious observer, experimenter, is involved into 
consideration of a measurement process. The latter was done by Wigner. He considered a physical system interacting with a measurement apparatus. The latter is under monitoring of an observer-experimenter whose nervous system detects and propagates the signal, e.g., visual, obtained from apparatus' pointer. By Wigner \cite{W1,W2} (see also \cite{Fred}), a measurement is completed by consciousness So, for him 
consciousness played the crucial role in the measurement process. The majority of the quantum community did not share Wigner's viewpoint; for them physical measurement is terminated at the pointer of an apparatus.} The brain, as a physical system, performs so to say self-measurements. To model such self-measurement, we split the brain, as an information processor,  into two subsystems, {\it unconsciousness} ${\cal UC}$ and {\it consciousness} ${\cal C}.$  The former plays the role of a system under observation and the latter  of an observer \cite{KHRfrontiers} (see \cite{p1}-\cite{pX} for mathematical modeling of join functioning of unconsciousness and unconsciousness  based on treelike geometry of the brain). To model the cooperation  of perceptions and emotions, the state space of  ${\cal UC}$  is decomposed into the tensor product of corresponding state spaces, 
$$
{\cal H}_{{\cal UC}} =   {\cal H}_{\rm{per}}\otimes {\cal H}_{\rm{em}}.
$$ 
Similarly, there are considered two classes of observables ${\cal A}$ and $O_{\rm{sup}},$ for conscious experiencing of perceptions and emotions, respectively.  

This paper is concentrated on modeling of emotional coloring of perceptions. But, we present the very general scheme of 
coloring of one class  of conscious experiences with another, ``basic experiences'' are colored with ``supplementary experiences''.
The aim and the origin of such coloring will be discussed below.  

Emotional coloring is modeled in the framework of quantum {\it contextuality} - emotional observables determine contexts for perceptions and other basic conscious experiences. We highlight that contextualization is a way  to reduce degeneration of unconscious states. Such contextual reduction is very important, since the state space of unconsciousness ${\cal H}_{{\cal UC}}$ has huge dimension and each conscious experience $x$ is based on multidimensional subspace ${\cal H}_x$ of  ${\cal H}_{{\cal UC}}.$ 

We stress that quantum-like modeling of brain's functioning should be sharply  distinguished from theories based on consideration of genuine quantum physical processes in the brain (cf. \cite{Liberman3} -- \cite{Igamberdiev2}). In the quantum-like approach, the brain is a macroscopic system  which information processing can be described by the mathematical formalism  of quantum theory 
(cf. Gunji et al. \cite{Gunji1,Gunji2}).  The quantum-like cognition project  (see, e.g., monographs \cite{UC3}, 
\cite{QL1} -- \cite{QL7}) does not contradict to the quantum cognition project. However, we proceed without the assumption that quantum features of information processing by the brain are coupled to quantum physical processes. The main distinguishing feature of the quantum information processing is operating with superpositions of alternatives. Such operating can be physically realized with classical 
electromagnetic waves. Genuine quantum information features are coupled to measurements  generating discrete events from superpositions.
\footnote{In Bohr's works \cite{BR}, such an event was called  phenomenon and its individuality was emphasized (see \cite{NL2} on comparison of quantum information processing in classical vs. quantum optics).}  

We treat brain's functioning in the  purely informational framework, so the states are not physical (electrochemical) states, but the information states. We do not need to localize     ${\cal UC}$ and ${\cal C}$ in the concrete  brain's areas. Of course, 
neuro-physiological studies give us coupling of ${\cal UC}$ and ${\cal C}$ to some areas in the brain, e.g., ${\cal C}$  to the
 prefrontal cortex. However, the distributed and physically nonlocalized information processing matches better our quantum-like model. We neither try to describe the neurophysiological mechanism of generation of perceptions and emotions and ``coloring''  of the 
former with the latter (see, e.g., \cite{LeDoux1} -- \cite{LeDoux3} for details). 

Of course, information processing in the brain is indivisibly coupled to electrochemical processes. However, as was emphasized by 
Liberman et al. \cite{Liberman1} -- \cite{Liberman6a} and Igamberdiev and Brenner \cite{Igamberdiev2a}, physical 
reduction of consciousness is impossible.\footnote{See, e.g., paper \cite{Liberman1}: {\it ``Biophysics cannot use the ordinary laws of physics and must take into account the influence on the phenomena to be studied, not only of a measurement but also of a calculation process in the real device predicting the future.} See, also  \cite{Liberman6a} {\it ``Living organisms measure many parameters in order to have orientation in the outer medium. That is why biophysics cannot use the ordinary laws of physics and must take into account the influence on the phenomena to be studied not only of a measurement but also of a calculation process in the real physical and biophysical device predicting the future.''}}

\section{The basic theories of consciousness}

As is well known, there are two basic competing theories of consciousness:
\begin{itemize}
\item the First Order Theory of   Consciousness \cite{FC1} -- \cite{FC5};
\item  the Higher Order Theory of   Consciousness \cite{SC1} -- \cite{SC3}.  
\end{itemize}

We characterize these theories with the following citation from article \cite{Paper}: 

``First-order theorists, such as Block, argue that processing related to a stimulus is all that is needed for there to be phenomenal consciousness of that stimulus \cite{FC1} -- \cite{FC5}. Conscious states, on these kinds of views, are states that make us aware of the external environment. Additional processes, such as attention, working memory, and metacognition, simply allow cognitive access to and introspection about the first-order state. In the case of visual stimuli, the first-order representation underlying phenomenal consciousness is usually said to involve the visual cortex, especially the secondary rather than primary visual cortex. Cortical circuits, especially involving the prefrontal and parietal cortex, simply make possible cognitive (introspective) access to the phenomenal experience occurring in the visual cortex.''

``In contrast, David Rosenthal and other higher-order theorists argue that a first-order state resulting from stimulus-processing alone is not enough to make possible the conscious experience of a stimulus. In addition to having a representation of the external stimulus one also must be aware of this stimulus representation. This is made possible by a HOR, which makes the first-order state conscious. In other words, consciousness exists by virtue of the relation between the first- and higher-order states. Cognitive processes, such as attention, working memory, and metacognition are key to the conscious experience of the first-order state. In neural terms, the areas of the GNC, such as the prefrontal and parietal cortex, make conscious the sensory information represented in the secondary visual cortex.''

The Higher Order Theory distinguishes between unconscious and conscious processing of mental information in the brain. By this theory, what makes cognition conscious is a higher-order  {\it observation} of the first-order processing. And in quantum theory observation is not simply inspection of the system's state. This is a complex process of interaction between a system and a measurement device.  This is the good place to cite Bohr \cite{BR}: 

\footnotesize{{\it ``This crucial point ...  implies the impossibility of any sharp separation between the behaviour of atomic objects and the interaction with the measuring instruments which serve to define the conditions under which the phenomena appear. In fact, the individuality of the typical quantum effects finds its proper expression in the circumstance that any attempt of subdividing the phenomena will demand a change in the experimental arrangement introducing new possibilities of interaction between objects and measuring instruments which in principle cannot be controlled.''}}

This viewpoint matches better with the Higher Order Theory of   Consciousness. A conscious experience is not simply introspection  of 
the ${\cal UC}$-state. 

In this paper, we do not describe the process of ${\cal UC}-{\cal C}$ interaction. In particular, we do not operate with the states 
of ${\cal C}.$ This can be done similarly to paper \cite{KHRfrontiers} by using the scheme of indirect quantum measurements in the framework of open quantum systems theory. However, it would make the presentation essentially more complicated from the mathematical viewpoint and 
the basic issue of this paper, contextuality, would be overshadowed by technicalities. We apply straightforwardly the canonical quantum observation theory as it was made by von Neumann \cite{VN}.\footnote{In \cite{Igamberdiev2} Igamberdiev considered another 
approach to quantum measurement theory, namely non-demolition quantum measurements, as the basis of bio-information processing. 
Such measurements differ from the von Neumann  measurements. In some sense, they are closer to classical measurements of electromagnetic waves. Although he referred to genuine physical processes, he emphasized that  in biosystems these processes are lifted to the macrolevel: {\it  ``The smallest details of living systems are molecular devices that realize non-demolition quantum measurements. These smaller devices form larger devices (macromolecular complexes), up to living body. The quantum device possesses its own potential internal quantum state (IQS), which is maintained for a prolonged time via reflective error-correction.}} 

The latter is slightly generalized by representation of observables by projector valued measures, instead of Hermitian operators. This generalization is motivated by two things.  Starting with a  Hermitian operator, it is possible to find its spectral decomposition and the projectors onto the subspaces corresponding to the eigenvalues, but the subspaces appear as secondary objects. In the POV representation of observers, subspaces are fundamental, and observer's values are associated with them. In our model, conscious experiences are associated with concrete subspaces of ${\cal H}_{{\cal UC}}.$ Another advantage of using POV-observables is that we are not obliged to label the values of observables by real numbers (operator's eigenvalues); any vocabulary can be used for description of conscious experiences.     

We remark that, although Bohr's viewpoint on the outcomes of quantum measurements dominates, a few respectable scientists 
claimed that these outcomes can be considered as the objective properties of physical systems.  Quantum measurements are treated as just approaching the premeasurement values of observables.
This position was presented in the  well know paper Einstein, Podolsky, and Rosen (EPR) \cite{EPR}; later Bell elaborated (and modified) EPR's argument \cite{Bell0,Bell1}, but he confronted the problem of nonlocality.  This line of thought matches better to the First Order Theory of   Consciousness. We shall follow the Bohr's line of thought \cite{BRR} and hence couple quantum measurement theory with the Higher Order Theory of   Consciousness. 

In application to emotions, the First Order Theory of Consciousness matches the somatic theories of emotions  rooted to James \cite{James}; the first of them is James-Lange theory \cite{JL}. Nowadays this viewpoint on emotions is advertized by some prominent 
neuro-physiologists, e.g. Damasio \cite{Damasio}.   

\section{Perceptions and emotions}

\subsection{Perception representation of sensations}

We follow to von Helmholtz \cite{Helmholtz} theory of sensation-perception. Perceptions are not simply a copies of sensations, not 
``impressions like the imprint of a key on wax'', but the results of complex signal processing including unconscious cognitive processing. In the modern science formulation, the process of creation of perception can be described as follows: \footnotesize{{\it ``Sensory information undergoes extensive associative elaboration and attentional modulation as it becomes incorporated into the texture of cognition. This process occurs along a core synaptic hierarchy which includes the primary sensory, upstream unimodal, downstream unimodal, heteromodal, paralimbic and limbic zones of the cerebral cortex.'' \cite{MES}}}
  
\subsection{Context-representation via emotions}

As is emphasized in \cite{Paper},  \footnotesize{{\it  ``Emotion schema are learned in childhood and used to categorize situations as one goes through life. As one becomes more emotionally experienced, the states become more differentiated: fright comes to be distinguished from startle, panic, dread, and anxiety.''}} In our terminology, each emotion-generation scheme is crystallized on of the basic life-contexts. {\it Context-labeling is the basic function of emotions.} Contextualization of surrounding environment was one of the first cognitive tasks of  biosystems and this ability was developed in parallel with establishing of sensation-perception system.

Memory is heavily involved in emotional activity, both for memorizing the features of contexts and for comparison of new perceptions with these contexts.  (See, e.g., \cite{LeDoux1} for the memory's role in cognition). Thus, evolutionary there was designed  a mental information processing system representing the basic life-contexts. This system is fixed at the level of the brain (and more generally the nervous system) hardware.
But, memorizing a variety of contextual experiences is done on the basis of the experiencing various situations (see above citation 
from \cite{Paper}).  This context-refection system was the root of the present emotion-system in humans. The latter has complex cognitive functions, not only contextual. However, in this paper we concern mainly contextuality.

Emotions represent adaptive reactions to environmental challenges; they are a result of human evolution;   
they provided optimal (from the viewpoint of computational resources) solutions to ancient and recurring problems that faced our ancestors \cite{Ekman}.\footnote{Although we do not follow the James-Lange theory of emotions \cite{JL}, this is the good place to mention that  James \cite{James} pointed out  that {\it ``feeling of the same changes as they occur is the emotion.} Here, for us the key words are {\it ``the same changes as they occur''}, i.e., the complex of repeatable bodily changes - physiological encoding  of a context.}

We emphasize that in our model emotions are conscious, cf. \cite{Paper}: \footnotesize{{\it  ``One implication of our view is that emotions can never be unconscious. Responses controlled by subcortical survival circuits that operate nonconsciously sometimes occur in conjunction with emotional feelings but are not emotions. An emotion is the conscious experience that occurs when you are aware that you are in particular kind of situation that you have come, through your experiences, to think of as a fearful situation. If you are not aware that you are afraid, you are not afraid; if you are not afraid, you aren't feeling fear.''}}

\section{Unconscious vs. conscious information processing in the brain}

\subsection{Unconsciousness}
\label{UC7}

An essential part of information processing in the brain is performed unconsciously; the information system responsible
for such processing (call it unconsciousness)  is denoted by the symbol ${\cal UC}.$ The space of its states is denoted by 
${\cal H}\equiv {\cal H}_{{\cal UC}}.$ In the quantum-like model, this is a complex Hilbert space (section \ref{QF}). 

The reader need not  couple  the notion of unconsciousness with the  names of  James \cite{J1}, Freud   \cite{F1}, and Jung \cite{Ju1}
(although  the author of this paper was strongly influenced by them, cf. with the previous works \cite{p1} -- \cite{pX}). In this paper, 
${\cal UC}$  denotes a special information processors of the brain.  It performs pre-observational  processing of the 
mental state. 

\subsection{Consciousness}

Perceptions and emotions are commonly treated as conscious entities. So, in our model the brain contains another 
information processing system generating conscious experiences; denote it by the symbol ${\cal C}.$
In our quantum-like framework, its functioning is modeled as performing measurements on 
the system ${\cal UC}.$ Introduction of two systems ${\cal UC}$ and ${\cal C}$ matches the quantum measurement scheme,
${\cal UC}$ is the analog  of a physical system exposed to measurements and ${\cal C}$ is the analog of a complex of 
measurement apparatuses. 

In the operational quantum approach, measurable quantities, observables, are represented 
by Hermitian operators acting in system's state space.  More advanced models within theory of open quantum systems
are based on consideration of the states of a measurement apparatus and interaction between the states of  the
system and apparatus \cite{DV} -- \cite{Oz1}. In our quantum-like modeling, the latter corresponds to consideration of the conscious states represented in complex Hilbert space  ${\cal H}_{{\cal C}}.$ In this paper we shall not consider (cf. 
\cite{KHRfrontiers,ENTROPY} for modeling the process of ${\cal UC}-{\cal C}$ interactions and generation of outputs of conscious observables).

Unconscious-conscious modeling of the brain's functioning matches well to the philosophic paradigm of the ontic-epistemic 
structuring of scientific theories. The ontic level is about reality (physical and mental) as it is if nobody observes it, 
the epistemic   level describes observations (see Atmanspacher \cite{AT1}). In contrast to quantum physics, in our model the epistemic 
level is related not to external observers, but to brain's self-observations. In the brain, unconscious and conscious processes are 
closely coupled and the sharp separation between them is impossible (cf. Brenner \cite{BRENNER} and Brenner and Igamberdiev
 \cite{BRENNER1}.

\subsection{Unconscious and conscious counterparts of the processes of generation of perceptions and emotions}
\label{EEE} 

In this paper, we shall be mainly concentrated on functioning of two information processors transforming 
\begin{itemize}
\item sensations $\to$ perceptions,
\item contexts $\to$ emotions.
\end{itemize}
Both processors have conscious outputs. Their functioning is strongly correlated; in the formalism  quantum theory   
correlations are represented by entangled states. We denote unconscious counterparts of these processors by the symbols
${\cal UC}_{\rm{per}}$ and  ${\cal UC}_{\rm{em}},$ respectively. In modeling of the emotional coloring of perception (its contextualization), we shall consider the compound information system $({\cal UC}_{\rm{per}},{\cal UC}_{\rm{em}}).$ 

This is the good place to mention the first theory of emotions, the James-Lange theory \cite{JL}. 
James claimed that {\it ``I am trembling. Therefore I am afraid.''} 
He stated: {\it ``My thesis ... is that the bodily changes follow directly the perception  of the exciting fact and that our feeling of the same changes as they occur is the emotion.''}  This way of thinking matches with the First Order Theory of   Consciousness and 
the EPR-Bell   viewpoint on quantum measurements. 

Following  LeDoux \cite{LeDoux1} (see also \cite{Paper}), we treat emotions within the Higher Order Theory of Consciousness coupled 
to Bohr's interpretation of quantum measurements.  

\subsection{Basic and supplementary conscious experiences, application to decision making}
\label{EEE1} 

Although we are mainly interested in emotional coloring of perceptions, the formalism under consideration can be applied to the very general class of compound information processing systems, $({\cal UC}_{\rm{bas}},{\cal UC}_{\rm{sup}}).$ The latter is used for determining stable repeatable and evolutionary fixed contexts for the former.  Simplest generalization of the perception-emotion scheme is emotional contextualization of decision making which modeling is based  
the compound system $({\cal UC}_{\rm{dm}},{\cal UC}_{\rm{em}}).$

Generation of conscious experiences (basic and supplementary) is modeled with POV-observables; denote the corresponding classes by the 
symbols $O_{\rm{bas}}$ and $O_{\rm{sup}}.$ In  particular, we shall consider the pairs $(O_{\rm{per}}, O_{\rm{em}})$ and 
$(O_{\rm{dm}}, O_{\rm{em}}).$  

An expert in cognition may suggest other pairs of basic and supplementary conscious experiences. One of the properties of the supplementary mental experiences is their rapid processing. They  should 
not inhibit processing of the basic mental experiences. We remark that emotion's  generation is characterized by high speed  
(see \cite{LeDoux1}).

\section{Quantum formalism: states and observables}
\label{QF}

Denote by ${\cal H}$ a complex Hilbert space. For simplicity, we assume that it is finite dimensional.  
Pure states of a system $S$ are given by normalized vectors of ${\cal H}.$ Later we shall consider mixed states
(section ), but, for the moment, we proceed with only pure states and call them simply statets.  

Physical observable $A$ is represented  by a Hermitian operator denoted by the same symbol.  Consider an operator with discrete spectrum; its spectral decomposition has the form:
\begin{equation}
\label{OP}
\hat A= \sum_x x\;  E^A_x,
\end{equation} 
where $E^A_x$ is the orthogonal projector onto the subspace of $H$ corresponding to the eigenvalue $x,$ i.e.,
${\cal H}^A_x=  E^A_x {\cal H}.$ We recall that the spectral family of orthogonal projectors satisfies the normalization 
condition 
\begin{equation}
\label{OPN}
\sum_x x\;  E^A_x =I,
\end{equation} 
where $I$ is the unit operator, and the mutual orthogonality condition
\begin{equation}
\label{OPO}
E^A_x \perp E^A_y, x \not= y,  
\end{equation} 
or equivalently 
\begin{equation}
\label{OPO1}
E^A_x E^A_y = \delta(x-y) E^A_x.  
\end{equation} 
The probability to get  the answer $x$ for a pure initial state $\psi$ is given by the Born rule
\begin{equation}
\label{BRRR1}
\Pr\{A =x\|\psi\} = \Vert  E^A_x \psi\Vert^2= \langle \psi\vert   E^A_x \psi \rangle .
\end{equation}
and according to the projection postulate (von Neumann \cite{VN})  the post-measurement state is generated  
by the map:  
\begin{equation}
\label{BRRR2}
\psi \to {\cal I}^A_x \psi = E^A_x \psi/ \Vert  E^A_x \psi\Vert.
\end{equation}
This state transformation is generated by observation's feedback to the system which initially was in the state $\psi,$ i.e., 
observations disturb systems' states. 

The quantum state update is the basis of quantum generalization of classical Bayesian inference . The projection update 
has properties which crucially differ  from the classical probability update. In particular, it generates 
violation of the law of total probability (playing the important role in Bayesian inference) \cite{KHC3,UC3,BU0} and  
the order effect which is absent in the classical probability theory (see Wang and Busemeyer \cite{WB},  Ozawa and Khrennikov \cite{ENTROPY} for its quantum-like modeling). The author of the present paper have stressed many times that non-Bayesian character of the quantum state and probability update is one of the distinguishing features of the quantum-like modeling of the brain's 
functioning \cite{KHRBa} (cf. Gunji et al. \cite{Gunji1,Gunji2}).

In physics, the values of observables are labeled by real numbers and it is convenient to represent observables 
by linear operators. But, in fact, the basic formulas (\ref{BRRR1}), (\ref{BRRR2}) contain only the orthogonal projectors $(E^A_x)$
and association of them with real numbers is not crucial. The label $x$ can belong to any set $X$ and then the probability
distribution given by Born's rule (\ref{BRRR1}) is defined on the set $X.$  The family of orthogonal projectors $(E^A_x)_{x \in X}$
encoding an observable $A$ should satisfy two constraints (\ref{OPN}), (\ref{OPO1}). They  guarantee that, for any state $\psi,$
quantity  $\Pr\{A =x\|\psi\}$ determined by Born's rule (\ref{BRRR1}) is a probability measure. (For simplicity, we consider only discrete sets of labels.) Such family of projectors  $(E^A_x)_{x \in X}$ is the simplest example of a {\it positive operator valued measure} (POVM), namely, the {\it projector-valued measure} (POV). The state transformation map (\ref{BRRR2}), $\psi \to {\cal I}^A_x \psi,$ determines the simplest 
{\it quantum instrument} \cite{DV}-- \cite{Oz1}. The symbol $A$ just labels an observable (not Hermitian operator), $A= (E^A_x).$ In principle, the same scheme can be realized with general quantum instruments and POVMs (see \cite{DV} -- \cite{Oz1}), but for simplicity we proceed  with POVs. 

\section{Incompatible conscious observables}
\label{CII}

In quantum physics, observables $C_1$ and $C_2$ are called compatible if they can be jointly measurable and the joint probability distribution (JPD) $p_\psi(C_1=x_1, C_2 =x_2)$ is well defined; observables which cannot be jointly measurable and, hence, 
their JPD cannot be defined  are called incompatible. In the mathematical formalism, compatibility and incompatibility are formalized through
 commutativity and noncommutativity, respectively. If observables are described as Hermitian operators $C_1,C_2,$ 
compatibility is encoded as $[C_1, C_2]=0;$ if they are POV-observables, then compatibility is encoded as 
\begin{equation}
\label{CI}      
 [E^{C_1}_{x_1}, E^{C_2}_{x_2}]= 0, \; \mbox{for all}\; x_1, x_2.
\end{equation}
Incompatibility is encoded as $[C_1, C_2]\not=0$ or, for POV-observables, as violation of (\ref{CI}) at least for one pair 
$(x,y).$  

For compatible observables, JPD is given by the following extension of the Born's rule:
\begin{equation}
\label{CI1}      
 p_\psi(C_1=x_1, C_2 =x_2)= \vert \langle E^{C_1}_{x_1} E^{C_2}_{x_2} \psi, \psi\rangle\vert^2 = \vert \langle E^{C_2}_{x_2} E^{C_1}_{x_1} \psi, \psi\rangle\vert^2.
\end{equation}
By using JPD compatible observables can be modeled in the classical probabilistic formalism. The formula (\ref{CI1})
can be generalized for an arbitrary number of observables $C_1,...,c_m,$ as 
\begin{equation}
\label{CI1}      
 p_\psi(C_1=x_1,..., C_m =x_m)= \vert \langle E^{C_1}_{x_1}.... E^{C_2}_{x_m} \psi, \psi\rangle\vert^2
\end{equation}
and this expression is invariant w.r.t. permutations.

We stress that the space of observables $O_{\rm{per}}$ can contain incompatible perceptions  as well as $O_{\rm{em}}$ can contain incompatible emotions.   In physics incompatibility is often seen as the exotic property of quantum theory - comparing with classical physical theory. Philosophically it is represented in the Bohr's complementarity principle which is difficult for understanding and 
was many times reformulated by Bohr  \cite{BR} .  
However, in the mental framework the notion of incompatibility can be interpreted very naturally:  there exist say emotions which can be experienced simultaneously; say happiness and sadness, pride and shame; in the same way it is evident that there exist incompatible, i.e., jointly unobservable perceptions and other conscious experiences. We note that mental observations are brain's self-observations. May be this self-observational property simplifies the incompatibility  issue.  

The necessity to operate with various incompatible entities is  the main roots of the use of the quantum(-like) information 
representation. In the absence of incompatibility, i.e., if, for the same mental state, the brain were able to construct the consistent probabilistic representation (in the form of JPD) of all possible combinations of say emotions, the quantum state formalism would be unnecessary.   

\section{Resolution of degeneration of states and  quantum contextuality}

Consider an observable $A$ which is mathematically described as POV $(E^A_x)_{x \in X}.$ For the discrete set of observance's values
$X=(x_k),$ we use notation $E_k^A= E_{x_k}^A.$
 Suppose now that some of its projections $E_k^A$ are degenerate; $\rm{dim}\;
{\cal H}_k^A >1,$ where ${\cal H}_k^A=  E_k^A {\cal H}.$  Moreover, for some outcomes, degeneration is very high, $\rm{dim}\;
{\cal H}_k^A >>1.$ In this case, a huge set of states  corresponds to the same outcome $x_k.$

Observer may be unsatisfied by such a situation; observer wants to refine his observation to split (at least partially) 
the states corresponding to the fixed outcome. How can it be done? The quantum measurement formalism presents the very natural and simple procedure for refinement of states' structure. 

Consider another quantum POV-observable $B = (E_m^B)$ compatible with the original observable $A,$ i.e.,
$[E_k^A, E_m^B]=0$ for all indexes $k, m.$  We remark that $B$-observable has its own set of outcomes, $Y=(y_m),$
which need not coincide with $X.$ 

For any (pure) quantum state $\psi,$ the POV-observables $A$ and $B$ can be jointly measurable with outcomes $(x_k,y_m)$
and the join probability distribution
\begin{equation}
\label{OOa}
p_{AB}(x_k, y_m| \psi)= \Vert E_k^A E_m^B \psi  \Vert^2= 
\Vert  E_m^B   E_k^A \psi  \Vert^2 .
\end{equation}
The projection postulate and commutativity of projectors imply that the post-measurement state is the same 
for the joint measurement with outcome  $(x_k,y_m)$ and sequential measurements, first $A=x_k$ and then 
$B=y_m$ or vice verse:
\begin{equation}
\label{BRRR2a}
\psi \to {\cal I}^A_x {\cal I}^B_y  \psi =  \frac{E^A_x E^B_y \psi}{ \Vert  E^A_x E^B_y \psi\Vert} =  \frac{ E^B_y E^A_x  \psi}{ \Vert  
E^B_y  E^A_x \psi\Vert} 
 =  {\cal I}^B_y  {\cal I}^A_x \psi.
\end{equation}
The state space ${\cal H}_k^A$ is reduced to the space ${\cal H}_{km}^{(A,B)}= E_k^A E_m^B {\cal H} (=  E_m^B E_k^A {\cal H});$
so 
\begin{equation}
\label{OO1}
{\cal H}_{k}^{A}= \oplus_m {\cal H}_{km}^{(A,B)}
\end{equation}
(direct sum decomposition).
Thus, the observer can specify its post-observation  state much better. 

The most fruitful refinement is based on an observable with nondegenerate spectra. Let all projectors  $ {\cal I}^B_m$
be one dimensional, $\rm{dim}\;{\cal H}_m^B = 1$ and let $e_m^B$ be the corresponding basis vector, i.e.,   
$ E_j^B  e_m^B = \delta(j-m) e_m^B.$ In this case, the outcome $(x_k,y_m)$ completely determines 
the post-observation mental state: $\psi \to e_m^B.$ 
The correspondence between labels $(x_k,y_m)$ and post-measurement states is one-to-one.

Consider now another observable $C$ which is also compatible with $A$, mathematically this is expressed as 
$[E^A_k, E^C_n]=0$ for all $k, n.$ The state spaces corresponding to $A$-outcomes can also be refined w.r.t. $C$-outcomes, i.e.,
\begin{equation}
\label{OO2}
{\cal H}_{k}^{A}= \oplus_m {\cal H}_{km}^{(A,C)},
\end{equation}
where  ${\cal H}_{km}^{(A,C)}= E_k^A E_m^C {\cal H} (=  E_m^C E_k^A {\cal H}).$ In particular, if, for all $n,$ 
$\rm{dim}\;{\cal H}_n^C = 1$ with the basis vector $e_m^B,$ i.e., $ E_j^C  e_n^C = \delta(j-m) e_n^C,$ then 
the outcome $(x_k,z_n)$  completely determines  the post-observation mental state: $\psi \to e_n^C.$

In quantum measurement theory, selection of observables co-measurable with $A$ is considered as specification of measurement context of $A$-measurements;
the $A$-value in the $B$-context can differ from  the $A$-value in the $C$-context, for the same premeasurment state $\psi.$ 
This is the essence of contextuality  playing so important role in quantum information theory \cite{BellX}:

{\bf Definition 1.} {\it If $A, B, C$ are three quantum observables, such that $A$ is compatible with $B$ and  $C,$ a measurement of $A$ might give different result depending upon whether $A$ is measured  with $B$ or with $C.$} 

We note that contextual behavior corresponds to the case of incompatible quantum observables $B$ and $C$, i.e., 
there exist indexes such that  $[E^B_m, E^C_n] \not=0.$ If all observables are pairwise commute, i.e., for all
indexes $[E^A_k, E^B_m]= [E^A_K, E^C_n] = [E^B_m, E^C_n]=0,$ then, for any state $\psi,$  it is possible to construct 
the noncontextual model of measurement based on the joint probability distribution for triple outcomes
\begin{equation}
\label{OO3}
p_{ABC}(x_k, b_m, c_n| \psi)= \Vert E_k^A E_m^B E_n^C \psi \Vert^2= ... =
\Vert E_k^A E_n^C  E_m^B E_n^C \psi \Vert^2 .
\end{equation}
If $B$ and $C$ are incompatible, such a model is impossible. This is the contextuality scenario. However, contextuality formalized
via Definition 1 cannot be tested experimentally, since it involves counterfactual reasoning. The only possibility is
test contextuality (based on Definition 1) indirectly with the aid of Bell-type inequalities (see appendix).

If the sets of observables' outcomes coincide with subsets of the real line, then the above considerations can be essentially simplified -
with the Hermitian linear operators representing observables. The contextuality scenario is related to observables 
satisfying the  commutation relations $[A, B]=[A,C]=0, [B, C]\not=0.$

However, the operator language is misleading and not only because representation of outcomes by real numbers is too special for coming 
cognitive applications. The main problem is that in the linear-operator approach the basic entities are outcomes, eigenvalues of the 
operator-observable. The subspace ${\cal H}^A_x$ is constructed as the space of the eigenvectors of the operator $A$
corresponding to the eigenvalue $x.$  In our coming modeling, we proceed another way around. The basic structures are mutually orthogonal subspaces ${\cal H}_k$ such that 
\begin{equation}
\label{OO7}
{\cal H}= \oplus_k {\cal H}_k.
\end {equation}
Then each of these subspaces is identified with some value of the POV-observable $A$ given by projectors on these subspaces. 
Roughly speaking, first states then values. 

\section{Resolution of degeneration of the states of consciousness via contextual coloring}

We start with the presentation of the general scheme for the resolution of degeneration of the ${\cal C}$-states. 
In this scheme ${\cal C},$ operates with two classes of observables $O_{\rm{bas}}$ and $O_{\rm{sup}}$ representing  {\it the ``basic and supplementary conscious  experiences''}, respectively. 

In the quantum-like model of generation of conscious  experiences developed by the author \cite{KHRfrontiers}, consciousness ${\cal C}$ is modeled 
as a system performing observations over unconsciousness ${\cal UC}.$  As in section \ref{UC7},  the 
symbol ${\cal H}$ denotes the space of ${\cal UC}$-states. A conscious observable $A$ is represented as a POV-observable $E^A_x, x \in X,$ where $X$ is the set of conscious 
experiences. The latter can be, for example,  the set of  language  expression or  visual images.
The values of an observable are determined by the subspaces ${\cal H}^A_x.$ If consciousness ${\cal C}$ detects a state 
belonging to ${\cal H}^A_x,$ then it generates the conscious experience $x \in X.$ 

The space of unconscious states  ${\cal H}$ has very high dimension. In reality, it is infinite dimensional, since this is the quantum information representation of 
electrochemical waves in the brain (see \cite{KHBR1} for details). We restrict modeling to the finite dimensional case for state spaces of high  dimension, $\rm{dim} \; {\cal H} >>1.$

 If the ``conscious-experience vocabulary'' $X$ (for observable A) is not so large, i.e., 
 number of points in set $X << \rm{dim} \; {\cal H},$ the same conscious experience $x$ is generated by huge variety of unconscious states. This degeneration is not good for cognitive behavior - reactions to external and internal stimuli and communications with 
other humans, especially for the latter. 

How can the brain, as the self-observable,  reduce this mental state degeneration? The answer is known from quantum theory. Consciousness ${\cal C}$ has to complete the $A$-observation, $A \in O_{\rm{bs}},$ with 
the observation of a compatible observable $B \in O_{\rm{sup}}.$ The latter plays the role of context for the $A$-observation. The value $A=x$ is contextualized 
with the value $B=y.$ In each outcome $(x,y)$ of the joint measurement of $(A, B),$  the value $x$ represents the basic conscious experience and $y$ its contextual coloring.   

The crucial point is that  such contextualization-observable $B$ should not carry  conscious meaning which is directly related to the $A$-meaning,  otherwise the joint observation $(A=x, B=y)$ can essentially modify the meaning of the outcome $A=x.$ It is also preferable that  $B$-observation can be combined not only with $A$-observation, but with observation of  any  $A^\prime \in  O_{\rm{bas}}.$

We proceed with POV-observables; in particular, to enrich selection of outcomes' vocabularies.
However, the reader use Hermitian operators and restrict all vocabularies 
to subsets of the real line. In the operator-observables framework, the condition of compatibility is formulated simply as
$[A, B]=0$ for any $A \in  O_{\rm{bas}}$ and any  $B \in O_{\rm{sup}}.$ In our POV-observables framework, we proceed with the condition
\begin{equation}
\label{OO8}
[E^A_x, E^B_y]=0\; \mbox{for observables}\;  A=(E^A_x), B=(E^B_y), 
\end{equation}
or symbolically
\begin{equation}
\label{OO8}
[O_{\rm{bas}}, O_{\rm{sup}}]=0.
\end{equation}
In general two observables $A, A^\prime \in O_{\rm{bas}}$  do not commute, i.e., there can exist two projectors such that 
$[E^{A}_x, E^{A^\prime}_{x^\prime}] \not=0.$ Thus, an arbitrary $A^\prime \in O_{\rm{bas}}$ cannot be used for refinement 
of $A \in O_{\rm{bas}}.$    

As was mentioned, for the outcome $A=x$ the co-outcome $B=y$ can be considered as coloring 
of the experience  $A=x.$ Cognitive $O_{\rm{bas}}$-representation can be compared with black-white pictures of houses in a town, the 
$O_{\rm{sup}}$-representation adds colors: the house $A=x$ is ``colored'' with the color $B=y.$ (Here we use ``coloring'' metaphorically.) Conscious experiences of $(x,y)$ and  $(x,\tilde{y}),$ where $x \in X$ and $y, \tilde{y} \in Y$
(the value-sets for $A$ and $B)$ can differ essentially. 

Thus, appeal to supplementary conscious experiences given by  the set of observables $O_{\rm{sup}}$ enriches tremendously the 
set of basic conscious experiences. At the level of mental states, it makes correspondence between 
states and experiences less degenerate. The latter helps a lot in social communication 
between individuals.
  
In principle,  $O_{\rm{bas}}$ also can contain compatible observables, say $A, A^\prime$ such that $[A,A^\prime] =0.$  
In such a case,  the outcomes of $A^\prime$ might be used by ${\cal C}$ for ``coloring'' of the outcomes of $A$ and vice verse. 
However, the $O_{\rm{sup}}$-coloring is preferable. The set of observables $O_{\rm{sup}}$ is specified by ${\cal C}$ at the level 
of hardware  and soft-ware; ${\cal C}$ need not to check whether an observable from $O_{\rm{sup}}$ can be used or not for coloring of an arbitrary 
observable $A \in O_{\rm{bas}}.$ Another problem with mutual coloring of  observables $A, A^\prime$ 
is that both carry important cognitive meanings  and coloring of $A$ by $A^\prime$ (or vice versa) can modify the cognitive meaning 
of $A,$ since ${\cal C}$ should process simultaneously two basic conscious functions.

 And finally, we repeat that generation of outcomes of observables belonging to $O_{\rm{bas}}$ is slower (in some situations, e.g., for emotional coloring, essentially slower \cite{LeDoux1}) than generation of $O_{\rm{sup}}$-observables. So,  by attempting to color $A \in O_{\rm{bas}}$ with $A^\prime \in O_{\rm{bas}}, \; {\cal C}$ would consume more time and even time scale inconsistency can be a problem.

\subsection{Emotional coloring of perceptions}

In the above scheme, we set $O_{\rm{bas}}\equiv O_{\rm{per}}$ and $O_{\rm{sup}}\equiv O_{\rm{em}}.$

As was stated in section \ref{EEE}, the information processing system ${\cal UC}$ contains the following  two subsystems. One is involved in  processing of sensations into perceptions -- still unconscious processing; denote it as ${\cal UC}_{\rm{per}}.$  Another system ${\cal UC}_{\rm{em}}$  processes unconscious emotional states.  The corresponding state spaces denote by the symbols ${\cal H}_{\rm{per}}\equiv {\cal H}_{{\cal UC}; \rm{per}}$ and ${\cal H}_{\rm{em}}\equiv {\cal H}_{{\cal UC}; \rm{em}},$ respectively.

Generally ${\cal UC}$   is not reduced to ${\cal UC}_{\rm{per}}$ and ${\cal UC}_{\rm{em}}.$  But, for simplicity, for modeling 
emotional coloring of perceptions,   
we assume that ${\cal UC}$ is compound solely of these subsystems.  This is really oversimplified picture of functioning of  unconsciousness which is not reduced to transformation of sensations into perceptions and generation of emotional states. But, we proceed in this framework.

\section{Tensor product formalism}

Now we turn again to the general scheme of modeling of basic and supplementary conscious experiences. Before
we have operated with the corresponding sets of observables $O_{\rm{bas}}$ and $O_{\rm{sup}},$ now we would like to proceed in the dual framework  by operating with  states. 

In quantum information theory, contextuality is typically presented within the tensor product formalism.\footnote{
In particular, it is closely coupled to the experimental tests of contextuality - with the Bell-type inequalities.
In fact, the direct experimental testing of contextuality in the sense of Definition 1 is impossible. The Bell test is about contextuality of possible hidden variables modedel beyond quantum observables. In physics, the contextuality issue is mixed with nonlocality, up to consideration of spooky action at a distance. Our quantum-like model does not reflect genuine quantum effects in the brain. Information processing is based on classical electromagnetic fields (cf. with applications of classical optics in quantum information theory \cite{NL2}). Contextuality is a consequence of existence of incompatible observables,
 in the spirit of articles \cite{NLa,NLb}.}

Let, as above, ${\cal H}$ be the state space of ${\cal UC}.$ Suppose that it is factorized in the tensor product 
${\cal H}= {\cal H}_{\rm{bas}} \otimes {\cal H}_{\rm{sup}},$ where ${\cal H}_{\rm{bas}}$ and ${\cal H}_{\rm{sup}}$ are state spaces coupled to basic 
and supplementary conscious experiences which are represented  at the conscious level by observables belonging to the
sets $O_{\rm{bas}}$ and  $O_{\rm{sup}}.$ These states spaces are generated by two different unconscious information processing 
systems, say ${\cal UC}_{\rm{bas}}$ and ${\cal UC}_{\rm{sup}}.$ These systems are concentrated in different  brain's areas created at the different stages 
of the brain's evolution. However, concentration is not sharp, processing is distributed and the processing areas have overlap.
The complex Hilbert space ${\cal H}$ is the state space of the compound system $({\cal UC}_{\rm{bas}},{\cal UC}_{\rm{sup}}).$ In the present model, it can be identified
with unconsciousness, ${\cal UC}.$ (Generally ${\cal UC}$ has more complex structure.)

Observables belonging to  $O_{\rm{bas}}$ and $O_{\rm{sup}}$ represent measurements which are performed by 
${\cal C}$  on the subsystems ${\cal UC}_{\rm{bas}}$ and ${\cal UC}_{\rm{sup}},$ respectively.  If observables are mathematically described as Hermitian operators, then
each operator $A:  {\cal H}_{\rm{bas}} \to {\cal H}_{\rm{bas}}$  is represented as  the operator 
${\bf A}= A\otimes I:  {\cal H} \to  {\cal H} $ and 
each operator $B: {\cal H}_{\rm{sup}} \to {\cal H}_{\rm{sup}}$  is represented as the operator 
${\bf B}= I \otimes  B: {\cal H} \to {\cal H}.$  For POV-observables,
we use the same procedure: for $A=(E^A_x)$ and $B=(E^B_y),$ we set  ${\bf A}= (E^A_x\otimes I)$ and 
${\bf B}= (I \otimes  E^B_y).$ The joint measurement of two POV-observables is represented 
by POV $A \otimes B =(E^A_x \otimes E^B_y).$ 

The state space ${\cal H}= {\cal H}_{\rm{bas}} \otimes {\cal H}_{\rm{sup}}$ is generated by tensor products of the form $\psi_{\rm{bas}} \otimes \psi_{\rm{sup}}.$  Measurements  of observables on such separable states  are reduced to independent measurements on the subsystems 
${\cal UC}_{\rm{bas}}$ and ${\cal UC}_{\rm{sup}}$ (for  $A,B$ of aforementioned classes). The real quantum information effects become visible for non-separable,
{\it entangled}, states, those states which cannot be represented in the form of the tensor product. 

For example, let both state spaces 
be two dimensional (qubit spaces), and let $\vert j_a\rangle$ and $\vert j_b\rangle, j_a, j_b=0,1,$ be orthonormal bases in 
${\cal H}_{\rm{bas}}$ and  ${\cal H}_{\rm{sup}},$ let observables $A$ and $B$ represented by one dimensional POV corresponding to these bases, i.e., $E^A_{j_a}=\vert j_a\rangle \langle j_a|$ and   $E^B_{j_b}=\vert j_b\rangle \langle j_b|,$ where $j_a, j_b=0,1.$ 
Set  $\vert j_a j_b \rangle= \vert j_a \rangle \otimes  \vert j_b \rangle;$ this is an orthonormal basis in the state space ${\cal H}$ 
of the compound system $S=({\cal UC}_{\rm{bas}},{\cal UC}_{\rm{sup}}).$ Then the states of $S$ can be expanded w.r.t. this basis. 
\begin{equation}
\label{L1}
\psi = \sum_{ji} c_{ji} \vert ji \rangle, \; \sum_{ji} \vert c_{ji}\vert^2=1. 
\end{equation}
By the Born's rule  $p_\psi(A=a_{j}, B=b_i) = \vert c_{ji}\vert^2$ is the probability that ${\cal C}$ observes the 
perception-emotion pair  $(a_{j}, b_i).$  

For example, the state   
\begin{equation}
\label{L2}
\psi = (\vert 00 \rangle +  \vert 11 \rangle)/\sqrt{2} 
\end{equation}
is entangled. This state illustrate the correlation meaning of entanglement - in fact, the maximal entanglement. 
Suppose that observables $A$ and $B$ yield the real values $a_{0}, a_{1}$ and $b_{0}, b_{1},$ respectively.
Consider quantum-like modeling of emotional coloring of perception, i.e., for ${\cal H}_{\rm{bas}}= {\cal H}_{\rm{per}}, 
{\cal H}_{\rm{sup}}= {\cal H}_{\rm{em}}.$ Perception $A=a_j$ is firmly associated with emotion $B=b_j, j=0,1.$ Thus, this state 
represents the perfect correlations between the $A$-perception and the $B$-emotion. 

For example, consider Russia or France in 19th century, decision maker was an officer who was participating in some social event
and conflicting with another officer; 
$A$ is the decision observable,  ``to challenge $(A=a_1)$ or not $(A=a_0)$ to a duel''; $B$ the emotion observable, 
``angry $(B=b_1)$ or not  $(B=b_0).$ If an officer is in the (unconscious) mental state (\ref{L2}), then the emotion ``angry''
matches perfectly with challenge to a duel.

The general two  qubit state given by  superposition (\ref{L1}) encodes correlation
\begin{equation}
\label{L3}
\langle AB \rangle_\psi \equiv \langle {\bf A} {\bf  B} \psi,\psi\rangle = \sum_{ji} a_{j} b_i \vert c_{ji}\vert^2 =  \sum_{ji} a_{j} b_i p_\psi(A=a_{j}, B=b_i),
\end{equation}
w.r.t. state $\psi \in {\cal H}.$
The last sum is the classical probabilistic expression for correlation. Thus, each concrete pair (perception, emotion) or (decision, emotion) can be described in the classical probabilistic framework. In particular, by experimenting with just one pair we would detect 
quantum-like effects. 

Consider, for the same perception $A \in O_{\rm{per}},$  another emotion $B^\prime \in  O_{\rm{em}}$ and 
correlation $\langle AB^\prime\rangle_\psi;$ it can be expressed with the coefficients with respect to the basic
$(\vert j_a j_{b^\prime} \rangle= \vert j_a \rangle \otimes  \vert j_{b^\prime} \rangle),$  
\begin{equation}
\label{L4}
\langle AB^\prime\rangle_\psi \equiv \langle {\bf  A} {\bf B}^\prime \psi,\psi\rangle = \sum_{ji} a_{j} b_i^\prime 
\vert c_{ji}^\prime\vert^2 =  \sum_{ji} a_{j} b_i^\prime p_\psi(A=a_{j}, B^\prime=b_i^\prime). 
\end{equation}
This is also the classical expression for the correlation.

\section{Bell type inequalities and experimental testing of emotional contextuality}

Now let us consider the correlation-expressions (\ref{L3}), (\ref{L4}) jointly. If emotional-observables $B=(E^B_y)$ and $B^\prime= (E^{B^\prime}_{y^\prime}$ are incompatible, i.e., in the mathematical terms, there 
exist projectors such that $[E^B_y, E^{B^\prime}_{y^\prime}]\not= 0,$
then generally it is impossible to combine these two classical correlations in the single classical probabilistic model. 
This is the complex foundational problem which is formalized with the aid of Bell type inequalities \cite{Bell0,Bell1,BellX}. We are not able to go deeper into basics of quantum mechanics. We finish  with the following foundational remark. In quantum physics, violation of 
these inequalities is coupled to violation of at least one of the followings two  assumptions:
\begin{itemize}
\item a) realism, i.e., the possibility to assign the values of observables before measurement; 
\item b) locality, the absence of action at a distance.
\end{itemize}
The combination of a)+b) is known as local realism. In quantum physics, one does not distinguish the two components of 
local realism (but, see \cite{NLa,NLb} for the claim that the key issue is violation of a) and that quantum theory is local; 
see also Plotnitsky \cite{PL1} -- \cite{PL3} for foundational analysis of the interplay incompatibility-nonlocality).  
As was argued in section \ref{CII}, ``mental realism'', i.e., the assumption that say all possible emotions peacefully 
coexist in any mental state, is hardly acceptable. Therefore we consider violation of ``mental realism'' as the root 
for violation of the Bell type inequalities.  

In quantum physics, experimental testing of the Bell type inequalities is the hot topic (see, e.g., \cite{E1,E2,E3} for the recent 
experiments). In psychology and decision making, they have been tested by a few authors \cite{CGB1} -- \cite{CGB5}. This paper can stimulate 
such experimenting in consciousness studies with joint measurements of the pairs $(A,B)=$ (perception, emotion) or 
(decision making, emotion). 

As in physics and the previous psychological experiments, it is natural to test the CHSH inequality
\cite{CHSH}. To proceed with it, it is not enough to consider just one perception (decision making) $A$ and two incompatible 
emotions $B$ and $B^\prime.$ One has to work as well with two incompatible perceptions $A$ and $A^\prime$ and form  
cyclically their correlations.The CHSH correlation function is given by the following combination of correlations:
\begin{equation}
\label{L5}
C_{\rm{CHSH}} = \langle AB \rangle + \langle AB^\prime \rangle +  \langle A^\prime B \rangle -
\langle A^\prime B^\prime \rangle 
\end{equation}
and, for dichotomous observables yielding values $\pm 1$  the following inequality holds:  
\begin{equation}
\label{L6}
\vert C_{\rm{CHSH}}\vert \leq 2.
\end{equation}

Finally, we stress that the psychological studies demonstrated that the humans behavior differs from the behavior of quantum physical systems w.r.t. so-called signaling problem \cite{CGB4,CGB5}.   It would be interesting to check whether experiments with emotions would lead to signaling  or not. My conjecture that, for some mental states, emotional coloring (contextualization) can be  performed without signaling. I stress that the proposed emotion-experiments differ from the previously performed psychological experiments in which 
experimenters operated with pairs composed of two decisions, i.e., one decision played  the role of context for another.

\section{Concluding remarks}

We presented the quantum-like model of emotional coloring of perceptions and other conscious experiences, including decision making.\footnote{We repeat that this model has no direct coupling to study of genuine quantum physical processes in the brain.}
The brain, as the information processor, is decomposed into two sub-processors, unconsciousness ${\cal UC}$ and 
consciousness ${\cal C}.$ The later plays the role of an observer on the former. This is mental realization of the quantum measurement
scheme for self-observatios performed by the brain. The state space of ${\cal UC}$ is mathematically described as a complex Hilbert space of very high dimension. In this paper we do not model the process of ${\cal UC}-{\cal C}$ interaction. Conscious observables 
are represented in the operator formalism. Perceptions and emotions are described by two classes of POV-observables, $O_{\rm{per}}$ and $O_{\rm{em}}.$ Perceptions are compatible with emotions, i.e., they can be jointly observed by ${\cal C}.$ In the mathematical formalism, compatibility is encoded as commutativity. 

Both perceptions and emotions are treated as conscious experiences. Quantum measurement formalism matches perfectly to the Higher
Order Theory of Consciousness. Emotions correspond to repeatable contexts and they contextualize perceptions and other conscious experiences. Emotional coloring reduces state degeneration for them and makes the information processing less diffuse. Context-matching is also important for social communication. 

One of the main distinguishing features of the quantum measurement theory is the presence of incompatible, i.e., jointly unobservable 
entities. In particular, the presence of incompatible observables makes impossible the use of the classical probability model 
(axiomatic of Kolmogorov \cite{K}).   The existence of incompatible perceptions or emotions is evident even from our personal experience. This motivates the use of the mathematical formalism fo quantum theory for modeling brain's self-observations.

To model emotional contextuality, we explore the tensor product formalism by factorizing the unconscious state space into the tensor product, one of its components describes the states of emotions. As is well known from quantum physics, the direct test of contextuality is impossible due to counterfactual nature of its formulation. The indirect tests of contextuality are based on the 
Bell type inequalities. We discuss the possibility of such tests for pairs (perception, emotion) or (decision making, emotion). 
As in physics, the problem of the interpretation of the violation of Bell inequalities  is very complex. Following \cite{NLa,NLb}, 
we couple violation and observables incompatibility, i.e., the existence of incompatible perceptions and emotions. 

We hope that our model can stimulate the applications of the quantum formalism in conscious-studies, especially in modeling perception-emotion and decision-emotion correlations.

\section{Appendix: Commutator representation of CHSH-correlation function } 

Here we present the purely mathematical result supporting our incompatibility interpretation of violation of the Bell type 
inequalities.

Consider Hermitian operators $A, A^\prime$ and $B, B^\prime.$ In our model, the first pair belongs to $O_{\rm{per}}$ and the second one 
to $O_{\rm{em}}.$ The considered perceptions and emotions are assumed to be compatible, i.e., 
$[A,B]=0, ..., [A^\prime, B^\prime]=0.$ It is also assumed that the observables are dichotomous and yield the values $\pm 1.$ In the operator terms, the latter  is expressed as $A^2= (A^\prime)^2= B^2= (B^\prime)^2=I,$ where $I$ is the unit operator.  

For quantum observables, the CHSH correlation function  $C_{\rm{CHSH}}$ can be represented as the average of the corresponding 
Hermitian operators (see \cite{NLb} for details): 
\begin{equation}
\label{C1}
C_{\rm{CHSH}}= \langle \Gamma_{\rm{CHSH}} \psi\vert \psi\rangle, 
\end{equation}
where
\begin{equation}
\label{C2}
\Gamma_{\rm{CHSH}}= A (B +  B^\prime) + A^\prime( B - B^\prime).
\end{equation}  
The CHSH inequality has the form: 
\begin{equation}
\label{C1a}
\vert \langle \Gamma_{\rm{CHSH}} \psi\vert \psi\rangle \vert \leq 2. 
\end{equation}
By straightforward calculation, one can derive at the Landau identity:
\begin{equation}
\label{C3}
\Gamma_{\rm{CHSH}}^2 = 4 I -  [A, A^\prime][B,B^\prime],
\end{equation}
where, for two operators $Q_1,Q_2, \; [Q_1,Q_2]=Q_1 Q_2 - Q_2 Q_1$ is their commutator.
Thus, if at least one pair of observables $(A,A^\prime)$ or $(B,B^\prime)$ is compatible, i.e.,
at least one of commutators $[A, A^\prime], [B,B^\prime]$ is equal to zero, then $\Gamma_{\rm{CHSH}} = 2 I$
and the CHSH inequality (\ref{C1a}) cannot be violated. If both commutators are nonzero, then, for some state $\psi,$
it can be violated (see \cite{}).     

The reader can see that this is purely commutativity-nocommutativity game.

\end{document}